\documentclass[12pt]{amsart}
\usepackage[T1]{fontenc}
\usepackage[english]{babel}
\usepackage{color}
\usepackage{hyperref}
\usepackage{dsfont}
\usepackage{mathabx}
\usepackage{enumitem}
\setlist[enumerate]{labelsep=*, leftmargin=1.5pc,
topsep=1ex plus0.5ex minus0.2ex,
itemsep=1ex plus0.5ex minus0.2ex,
font=\rmfamily,
font=\upshape}
\newtheorem{thm}{Theorem}[section]
\newtheorem{cor}[thm]{Corollary}
\newtheorem{lem}[thm]{Lemma}
\newtheorem{fac}[thm]{Fact}
\theoremstyle{definition}
\newtheorem{rem}[thm]{Remark}
\newtheorem{exa}[thm]{Example}
\numberwithin{equation}{section}
\newcommand{\id}{\mathds{1}}
\newcommand{\tr}{\operatorname{tr}}
\newcommand{\rk}{\operatorname{rank}}
\newcommand{\ii}{\operatorname{i}}
\newcommand{\cM}{\mathcal M}
\newcommand{\bE}{\mathbb E}
\newcommand{\bC}{\mathbb C}
\newcommand{\bN}{\mathbb N}
\newcommand{\bP}{\mathbb P}
\newcommand{\bR}{\mathbb R}
\begin{document}
\selectlanguage{english}
\thispagestyle{empty}
\pagestyle{myheadings}
\author{Stephan Weis}
\title{Maximum-entropy inference and
inverse continuity of the numerical range}
\begin{abstract}
We study the continuity of the maximum-entropy inference map 
for two observables in finite dimensions. We prove that the 
continuity is equivalent to the strong continuity of the 
set-valued inverse numerical range map. This gives a 
continuity condition in terms of analytic eigenvalue functions 
which implies that discontinuities are very rare. It shows 
also that the continuity of the MaxEnt inference method is 
independent of the prior state.
\end{abstract}
\vspace{.3in}

\maketitle
\markleft{\hfill Maximum-entropy inference and inverse continuity\hfill}
\markright{\hfill S.~Weis \hfill}

Key Words: maximum-entropy inference, continuity,
numerical range, strong continuity, stability, strong stability.
\vspace{.2in}

2000 Mathematics Subject Classification: 
Primary 81P16, 62F30, 94A17, 54C10, 47A12, 54C08; 
Secondary 47N50, 82B26.
\vspace{.3in}
%
%
%
%
%
%
\section{Introduction}
\par
The maximum-entropy principle, going back to Boltzmann, 
is one of the standard techniques in quantum mechanical 
inference problems 
\cite{Jaynes1957,Wichmann1963,Ingarden-etal1997,Streater2011,Ali-etal2012} 
and state reconstruction \cite{Buzek-etal1999,SwingleKim2014}. 
Here we consider a finite set of quantum observables, represented 
by hermitian matrices in the algebra $M_d$ of complex 
$d\times d$-matrices, $d\in\bN$. If their expected 
values with respect to several quantum states are identical 
then no unique quantum state is specified by these expected values.
The maximum-entropy inference map makes a definite choice by
selecting the state with maximal von Neumann entropy. This
inference map from expected values to states
can have discontinuity points on the boundary of 
the set of expected values \cite{WK,Weis-cont} while analogous 
inference maps to probability distributions are always continuous. 
The discontinuities have a meaning in physics. They have been 
discussed as a signature of a quantum phase transition 
\cite{Chen-etal2014}. They are passed 
\cite{Weis-GSI14,Rodman-etal2016} 
from the inference map to a correlation quantity 
called irreducible correlation 
\cite{Linden-etal2002,Zhou2008} which is connected to the topological 
entanglement entropy used to characterize topological order 
\cite{Liu-etal2014,Kato-etal2015}.
\par
Methods to analyze the discontinuities have included information 
topology 
\cite{Weis-topo}, convex geometry \cite{Weis-cont,Rodman-etal2016}, 
and, for two observables, numerical range techniques 
\cite{Rodman-etal2016}. Here, we focus on the case of two observables
which we encode into a single matrix $A\in M_d$ as its {\em real part} 
$\Re(A):=\tfrac{1}{2}(A+A^*)$ and {\em imaginary part} 
$\Im(A):=\tfrac{1}{2\ii}(A-A^*)$, a notation which we will meet again 
in Sec.~\ref{sec:rarity}. The set of {\em density matrices} in $M_d$ 
is denoted by
\[\textstyle
\cM_d:=\{\rho\in M_d\mid\rho\succeq 0,\tr(\rho)=1\}.
\]
This set is also called {\em state space} \cite{Alfsen-Shultz}, 
$a\succeq 0$ means that the matrix $a\in M_d$ is positive 
semi-definite. The state space is a convex body, that is a 
compact convex subset in a Euclidean space. The inner product 
$\langle a,b\rangle:=\tr(a^*b)$, $a,b\in M_d$, and norm 
$\|a\|_2:=\sqrt{\langle a,a\rangle}$ shall be used. 
\par
In quantum 
mechanics, see for example \cite{BZ}, Secs.~5.1 and~5.2, 
elements of $\cM_d$ represent {\em states} of a quantum
system and the real number $\tr(\rho a)=\langle\rho,a\rangle$, for 
an observable $a\in M_d$ and for $\rho\in\cM_d$, is interpreted 
as the expected value of $a$ when the system is in the 
state $\rho$. Since we use the map $\rho\mapsto\langle\rho,A\rangle$ 
in various restrictions the notation will simplify by reserving a 
symbol it. We define the {\em expected value functional}
\[\textstyle
\bE_A: \quad \{b\in M_d\mid b^*=b\}\to\bC, \quad
a\mapsto\langle a,A\rangle
\]
on the Euclidean space of hermitian matrices. 
The map $\bE_A$ sends a state $\rho\in\cM_d$ to the pair
$\bE_A(\rho)=(\langle\rho,\Re(A)\rangle,\langle\rho,\Im(A)\rangle)$
of expected values of the observables $\Re(A)$ and $\Im(A)$,
in the identification of the range $\bC$ with $\bR^2$.
\par
The domain of the maximum-entropy inference map is the convex
body
\[\textstyle
L_A:=\{\bE_A(\rho)\mid\rho\in\cM_d\}\subset\bR^2,
\]
comprising the expected value pairs of $\Re(A)$ and $\Im(A)$.
We call $L_A$ {\em convex support} \cite{WK,Weis-cont,Rodman-etal2016} 
by its name in probability theory \cite{Barndorff}. The 
{\em von Neumann entropy} of a state $\rho\in\cM_d$ is
$S(\rho)=-{\rm tr}(\rho\cdot\log\rho)$ and the 
{\em maximum-entropy inference} is the map
\[
\rho_A^*:\quad L_A\to\cM_d,\quad
\alpha\mapsto{\rm argmax}\{S(\rho)\mid\rho\in\cM_d,\bE_A(\rho)=\alpha\}.
\]
See \cite{Jaynes1957,Ingarden-etal1997} for more information about 
$\rho_A^*$. Our analysis will be based on \cite{Weis-cont}, Thm.~4.9, 
which affirms that for all $\alpha\in L_A$ 
\begin{equation}\label{eqiv-cont-open}
\text{$\rho_A^*$ is continuous at $\alpha$ if, and only if, 
$\bE_A|_{\cM_d}$ is open at $\rho_A^*(\alpha)$.}
\end{equation} 
Thereby, a function between topological spaces is {\em open at} a 
point in the domain if the image of every neighborhood of that point 
is a neighborhood of the image point. Clearly, every linear map is 
open in finite dimensions but it may fail to be open when restricted.
\par
Exact bounds on the number of discontinuity points of $\rho_A^*$ are 
known for $d\leq5$, see Secs.~7 and~8 in \cite{Rodman-etal2016}. The 
bounds have been derived from pre-image results \cite{Leake-etal-OAM2014}
of the following map $f_A$. The aim of this article 
is to go beyond these pre-image results and to establish a direct link 
to a continuity problem in operator theory
\cite{Corey-etal2013,Leake-etal-LMA2014,LinsParihar2015}. Denoting by 
$S\bC^d$ the unit sphere of $\bC^d$, the {\em numerical range map} of a 
matrix $A\in M_d$ is defined by
\[\textstyle
f_A: \quad S\bC^d\to\bC, \quad
x \mapsto\langle x,Ax\rangle.
\]
The {\em numerical range} is the image $W(A)=f_A(S\bC^d)$. Here, 
$\langle x,y\rangle:=\overline{x_1}y_1+\cdots+\overline{x_d}y_d$,
$x,y\in\bC^d$, is the inner product of $\bC^d$. The numerical range 
\cite{Kippenhahn1951,HornJohnson1991}
is a convex set by the Toeplitz-Hausdorff theorem \cite{LiPoon2000}
and it is well-known that $W(A)=L_A$ holds, see for example 
\cite{BerberianOrland1967}, Thm.~3. The set-valued inverse 
$f_A^{-1}:W(A)\to S\bC^d$ is called 
\cite{Corey-etal2013,Leake-etal-LMA2014,LinsParihar2015}
{\em strongly continuous} at $\alpha\in W(A)$ if for all 
$x\in f_A^{-1}(\alpha)$ the map $f_A$ is open at $x$. 
\par
Our main result can be summarized as follows.
\begin{thm}\label{thm1}
For all $\alpha\in L_A$ the maximum-entropy inference map 
$\rho_A^*$ is continuous at $\alpha$ if and only if 
$f_A^{-1}$ is strongly continuous at $\alpha$.
\end{thm}
\par
We remark that there is a very large set of functions which can 
replace the von Neumann entropy in this continuity analysis. Although 
the map $\rho_A^*$ will change, its topological properties will remain 
if $\rho_A^*(\alpha)$ lies in the relative interior of the fiber 
$\bE_A^{-1}(\alpha)$ for all $\alpha\in L_A$ (see Sec.~\ref{sec:prior}
for examples). This is the content of
Coro.~\ref{cor:major} but remains an open problem for more than two 
observables. Otherwise, if some of the inference points belong to 
the relative boundary of fibers, the topology can change already 
for two observables, see \cite{Rodman-etal2016}, Thm.~7.1.
We recall that the relative interior of a convex set $C$ is the 
interior of $C$ in the topology of the affine hull of $C$. The 
relative boundary of $C$ is the complement of the relative interior 
in the closure of $C$. 
\par
The proof of Thm.~\ref{thm1} at the end of 
Sec.~\ref{sec:MaxEnt-to-MinEnt} uses two properties of 
convex sets: Firstly (see Sec.~\ref{sec:MinEnt-to-MaxEnt}), 
like all two-dimensional convex bodies, the set $L_A$ is 
{\em stable}, that is the mid-point map 
$(\alpha,\beta)\mapsto\tfrac{1}{2}(\alpha+\beta)$ is open 
\cite{Papadopoulou1977}. Secondly (see 
Sec.~\ref{sec:MaxEnt-to-MinEnt}), the state space $\cM_d$ is 
{\em strongly stable} \cite{Shirokov2010}. These two notions 
of stability are equivalent for a large class of convex sets
\cite{Shirokov2012}. Stability of the set of 
density matrices on a separable Hilbert space was used, 
for example, to show the continuity of entanglement monotones 
\cite{ProtasovShirokov2009} arising from the 
convex roof extension \cite{Uhlmann2010}. The present work shows
that already in finite dimensions the topology of a linear map 
on the state space $\cM_d$ and the stability of its linear
images (see last paragraph of Sec.~\ref{sec:overview})
have much more to explore.\\
\par\normalsize
{\par\noindent\footnotesize
{\em Acknowledgements.}
Thanks to Lia L.~Pugliese for her help proofreading the English version 
of this essay, to Maksim E.~Shirokov for his help with the literature about 
stability, and to Ilya M.~Spitkovsky for his hospitality at the New York 
University Abu Dhabi (12/2014) and for discussions about eigenvalue 
functions. Special thanks to Arleta Szko{\l}a and Nihat Ay for general
support at the MPI for Mathematics in the Sciences in Leipzig and to the 
local library there, where I have enjoyed a pleasant working environment
(10/2014--05/2015).}
%
%
\section{Remarks and Corollaries}
\label{sec:overview}
\par
We comment on the main result, derive some corollaries, and 
provide an outlook.
\par
Thm.~\ref{thm1} is surprising because the functions $\rho_A^*$ and 
$f_A^{-1}$ are opposite in several respects. 
\begin{rem}\label{rem1}~
\begin{enumerate}
\item
Studying the continuity of $\rho_A^*$ requires by (\ref{eqiv-cont-open}) 
to check the openness of $\bE_A$ restricted to the state space $\cM_d$, 
while studying the strong continuity of $f_A^{-1}$ requires by 
Lemma~\ref{lem:cont-m-m1-cs} to check the openness of $\bE_A$ restricted 
to the extremal points $\cM_d^1$ of $\cM_d$.
\item
Lemma~5.8 in \cite{Weis-cont} shows that for all $\alpha\in L_A$ 
the state $\rho_A^*(\alpha)$ lies in the relative 
interior of the fiber $F:=\bE_A|_{\cM_d}^{-1}(\alpha)$. On the other 
hand, the set $f_A^{-1}(\alpha)$ consists of extremal points of $F$. 
In fact, we have seen in (1) that the elements of
$f_A^{-1}(\alpha)$ are extremal points of the state space $\cM_d$.
{\em A fortiori} they are extremal points of $F$.
\item
While for all $\alpha\in L_A$ the state $\rho_A^*(\alpha)$ maximizes 
the von Neumann entropy on the fiber $\bE_A|_{\cM_d}^{-1}(\alpha)$, 
the pre-image $f_A^{-1}(\alpha)$ is the zero level set of minimal 
entropy, see for example \cite{Wehrl1978}, Sec.~A.2.
\end{enumerate}
\end{rem}
\par
Two corollaries are worth pointing out. Sec.~\ref{sec:rarity} focusses 
on a continuity condition of $f_A^{-1}$ in terms of analytic eigenvalue 
functions
\cite{Leake-etal-LMA2014}. The condition then governs the continuity
of the inference map $\rho_A^*$. For example, this shows that $\rho_A^*$ 
has at most finitely many points of discontinuity and that the set of 
matrices $A$ where $\rho_A^*$ is continuous is open and dense in $M_d$. 
\par
Sec.~\ref{sec:prior} addresses the quantum {\em MaxEnt} inference 
method \cite{ShoreJohnson1980,Skilling89,CatichaGiffin2006,Ali-etal2012} 
which is an updating rule from a prior state to an inference state, 
given new information in terms of expected values. The maximum-entropy 
inference is the special case of a uniform prior. We prove that 
the MaxEnt inference, seen as a function from expected values to 
inference states, has for all prior states the same points of 
discontinuity.
\par
Finally, we remark that new ideas will be needed to extend the methods
of this article to $r=3$ (or more) observables $u_1,\ldots,u_r\in M_d$. 
Firstly, it was observed in \cite{Chen-etal2014}, Exa.~6 (see also
\cite{Rodman-etal2016}, Exa.~5.2) that the convex 
support $\{(\langle\rho,u_1\rangle,\ldots,\langle\rho,u_r\rangle)
\mid\rho\in\cM_d\}$ is not stable for some choices of observables 
$u_1,u_2,u_3\in M_3$ when $r=d=3$. Secondly, although the joint 
numerical range $\{\langle x,u_ix\rangle_{i=1}^r\mid x\in S\bC^d\}$ 
contains the extremal points of the convex support, it is in general 
not convex \cite{LiPoon2000} for $r\geq 3$. So the boundary of the 
convex support will need a careful analysis when trying to go 
beyond $r=2$.
%
%
%
\section{Preliminaries}
\label{sec:prelim}
\par
We introduce faces of convex sets and pure states.
We connect the domains of the functions $f_A$ and $\bE_A|_{\cM_d}$ 
by recalling properties of the quotient map $\beta:S\bC^d\to\bP\bC^d$
from the unit sphere in $\bC^d$ to the projective space of lines in 
$\bC^d$. This is a well-known smooth (even real analytic) map.
Nevertheless we provide a proof because we are also interested in
the openness of $\beta$.
\par
A {\em face} of a convex set $C$ 
is a convex subset $F\subset C$ such that if for $x,y\in C$ the open 
segment $]x,y[\,:=\{(1-\lambda)x+\lambda y\mid\lambda\in(0,1)\}$
intersects $F$, then the closed segment 
$[x,y]:=\{(1-\lambda)x+\lambda y\mid\lambda\in[0,1]\}$ belongs to 
$F$. An {\em extremal point} is a face of dimension zero and a
{\em facet} is a face of dimension $\dim(C)-1$.
\par
The extremal points of the state space $\cM_d$, $d\in\bN$, are called 
{\em pure states} in 
physics \cite{BZ,NC} and it is well-known, see for example (4.2) in 
\cite{Alfsen-Shultz}, that the set of pure states equals the set of 
rank-one density matrices which we denote by
\begin{equation}\label{eq:pure}\textstyle
\cM^1_d:=\{\rho\in\cM_d\mid\rk(\rho)=1\}.
\end{equation}
The rank-one density matrices are the orthogonal 
projections onto one-dimensional subspaces of $\bC^d$. So 
$\cM_d^1\cong\bP\bC^d=S\bC^d/S\bC^1$ is a projective space. 
We denote the quotient map in Dirac's notation
\begin{equation}\label{eq:beta}\textstyle
\beta: \quad S\bC^d\to\cM_d^1,\quad x\mapsto |x\rangle\langle x|.
\end{equation}
Its fibers are isomorphic to the circle $S\bC^1$. For $d=2$ the 
famous Hopf fibration is obtained, see for example \cite{BZ}.
\par
In the following we use the trace distance and
the fidelity, see for example \cite{NC}, Sec.~9.2.1--2. Let 
$\sqrt{a}$ denote the square root of a positive 
semi-definite matrix $a\in M_d$, that is $\sqrt{a}\succeq 0$ and 
$(\sqrt{a})^2=a$. The {\em trace norm} of $a\in M_d$ is 
$\|a\|_1:=\tr\sqrt{a^*a}$. The {\em trace distance} between 
states $\rho,\sigma\in\cM_d$ is
\[\textstyle
D(\rho,\sigma):=\tfrac{1}{2}\|\rho-\sigma\|_1
\]
and their {\em fidelity} is
$F(\rho,\sigma):=\|\sqrt{\rho}\,\sqrt{\sigma}\|_1
=\tr\,\sqrt{\sqrt{\rho}\,\sigma\sqrt{\rho}}$.
\par
The fidelity is symmetric in the two arguments by Uhlmann's theorem
\cite{Uhlmann1976}. We have $0\leq F(\rho,\sigma)\leq 1$ where the
upper bound is achieved if and only if $\rho=\sigma$. The
Fuchs-van de Graaf inequalities \cite{FuchsVanDeGraaf1999} 
\begin{equation}\label{inequFD}\textstyle
1-F(\rho,\sigma)\leq D(\rho,\sigma)\leq\sqrt{1-F(\rho,\sigma)^2}
\end{equation}
hold. For pure states we have 
$F(|x\rangle\langle x|,|y\rangle\langle y|)=|\langle x,y\rangle|$,
$x,y\in S\bC^d$, where $|z|:=\sqrt{\langle z,z\rangle}$ is the norm
of $z\in\bC^d$.
\par
We say a function between topological spaces is {\em open on} a 
subset of its domain if it is open at each point of this subset. 
The function is {\em open} if it is open on the domain.
\begin{lem}\label{lem:beta}
The map $\beta:S\bC^d\to\cM_d^1$ is continuous and open.
\end{lem}
{\em Proof:}
The second inequality in (\ref{inequFD}) shows
$D^2\leq (1-F^2)\leq 2(1-F)$. So, for $x,y\in S\bC^d$ we have
\begin{align*}\textstyle
D(|x\rangle\langle x|,|y\rangle\langle y|)^2 & \leq
|x|^2 +|y|^2-2|\langle x,y\rangle|
\leq |x|^2+|y|^2-2\Re(\langle x,y\rangle) \\
 & = |x-y|^2,
\end{align*}
whence $\beta$ is Lipschitz-continuous with the global constant 
one. The left-hand side inequality in (\ref{inequFD}) implies 
for all $x,y\in S\bC^d$ such that $\langle x,y\rangle\geq 0$ the 
inequality of
\[\textstyle
|x-y|^2
= 2(1-|\langle x,y\rangle|)
= 2(1-F(|x\rangle\langle x|,|y\rangle\langle y|))
\leq 2D(|x\rangle\langle x|,|y\rangle\langle y|).
\]
This proves that the ball in $S\bC^d$ of (Hilbert space) radius $\epsilon>0$
about $x\in S\bC^d$, mapped through $\beta$, contains the ball in $\cM_d^1$ 
of (trace distance) radius $\tfrac{1}{2}\epsilon^2$ about 
$|x\rangle\langle x|$. Hence $\beta$ is open.
\hspace*{\fill}$\square$\\
\par
Turning to the convex support and to the numerical range we notice
that for all $x\in S\bC^d$
\begin{equation}\label{eq:f=Eb}\textstyle
f_A(x)=\langle x,Ax\rangle
=\tr(|x\rangle\langle x|A)=\bE_A(|x\rangle\langle x|)=\bE_A\circ\beta(x).
\end{equation}
\begin{lem}\label{lem:cont-m-m1-cs}
For all $x\in S\bC^d$ the following statements are equivalent.
\begin{enumerate}
\item The map $f_A$ is open at $x$.
\item The map $\bE_A|_{\cM_d^1}$ is open at $|x\rangle\langle x|$.
\end{enumerate}
\end{lem}
{\em Proof:}
Using (\ref{eq:f=Eb}), {\rm  (1)}$\implies${\rm  (2)} follows from 
the continuity of $\beta$ and {\rm  (2)}$\implies${\rm  (1)} follows 
from the openness, proved in Lemma~\ref{lem:beta}.
\hspace*{\fill}$\square$\\
%
%
%
%
%
%
\section{Strong Continuity Implies Continuity}
\label{sec:MinEnt-to-MaxEnt}
\par
We prove that the strong continuity of $f_A^{-1}$ implies the
continuity of the maximum entropy inference $\rho_A^*$. The main argument 
is the stability of two-dimensional convex bodies.
\par
A convex body $C$ is {\em stable} if $C\times C\to C$, 
$(x,y)\mapsto\tfrac{1}{2}(x+y)$ is an open map
\cite{Papadopoulou1977,ClausingPapadopoulou1978}. We recall two 
facts about stable convex bodies. Firstly, if $C$ is a stable 
convex body then for any integer $n\geq 2$ and for
$(\lambda_1,\ldots,\lambda_n)\in\bR^n$ such that $\lambda_i\geq0$ for 
$i=1,\ldots,n$ and $\lambda_1+\cdots+\lambda_n=1$ the map
\begin{equation}\label{eq:barycenter}\textstyle
\underbrace{C\times \cdots \times C}_{\text{$n$ times}}\to C, \quad
(x_1,\ldots,x_n)\mapsto \lambda_1x_1+\cdots+\lambda_nx_n
\end{equation}
is open. The proof that (\ref{eq:barycenter}) is open is given for 
$n=2$ in \cite{ClausingPapadopoulou1978}, Prop.~1.1, and the 
case of $n\geq 3$ follows by induction. 
\par
Secondly, every two-dimensional convex body is stable. This follows 
from Thm.~2.3 in \cite{Papadopoulou1977} which says that a convex 
body $C$ of any finite dimension $l\in\bN$ is stable if, and only if, 
for each $k=0,\ldots,l$ the {\em $k$-skeleton}, that is the union 
of all faces of $C$ of dimension at most $k$, is closed. The 
$(l-2)$-, the $(l-1)$- and the $l$-skeletons of $C$ are always 
closed, see \cite{Grzaslewicz}, so every two-dimensional convex body 
is stable.
\par
The Euclidean ball of radius $\epsilon>0$ about $\rho\in\cM_d$ within 
a subset $C\subset\cM_d$ will be denoted by
$B_\epsilon(\rho,C):=\{\sigma\in C\mid\|\rho-\sigma\|_2\leq \epsilon\}$.
\begin{thm}\label{thm:min-to-max}
Let $\alpha$ be an extremal point of $L_A$.
If $f_A^{-1}$ is strongly continuous at $\alpha$
then $\bE_A|_{\cM_d}$ is open on 
$\bE_A|_{\cM_d}^{-1}(\alpha)$.
\end{thm}
{\em Proof:}
If $\alpha\in L_A$ is an extremal point then the fiber
$F:=\bE_A|_{\cM_d}^{-1}(\alpha)$ is a face of the state space $\cM_d$, 
so all extremal points of $F$ are pure states or equivalently, by
(\ref{eq:pure}), they belong to the set of rank-one states $\cM_d^1$. 
Hence we can write an arbitrary point $\rho\in F$ in the form
\[\textstyle
\rho=\lambda_1\rho_1+\cdots+\lambda_n\rho_n
\]
where $\rho_i\in\cM_d^1\cap F$, $\lambda_i\geq 0$ for $i=1,\ldots,n$ 
and $\lambda_1+\cdots+\lambda_n=1$. Let $x_i\in S\bC^d$ such that 
$\rho_i=|x_i\rangle\langle x_i|$ and choose a neighborhood 
$N_i\subset\cM_d^1$ 
of $\rho_i$ in $\cM_d^1$. By the continuity of $\beta:S\bC^d\to\cM_d^1$
(see Lemma~\ref{lem:beta}) the pre-image 
$N'_i:=\beta^{-1}(N_i)$ is a neighborhood of $x_i$ in $S\bC^d$. The 
assumption that $f_A^{-1}$ is strongly continuous at 
$\alpha$ proves that $f_A(N'_i)$ is a neighborhood of $\alpha$. Hence 
$\bE_A(N_i)=f_A(N'_i)$ is a neighborhood of $\alpha$.
\par
Now let $N\subset\cM_d$ be an arbitrary neighborhood of $\rho$ in $\cM_d$ 
and choose neighborhoods $N_i\subset\cM_d^1$ about $\rho_i$ in $\cM_d^1$
such that $\lambda_1N_1+\cdots+\lambda_nN_n\subset N$. It suffices to
consider a Euclidean ball $B_\epsilon(\rho,\cM_d)\subset N$ of radius 
$\epsilon>0$ about $\rho$ and to use the Euclidean balls 
$N_i=B_\epsilon(\rho_i,\cM_d^1)$ about $\rho_i$, $i=1,\ldots,n$. Then
\[\textstyle
\lambda_1\bE_A(N_1)+\cdots+\lambda_n\bE_A(N_n)\subset \bE_A(N).
\]
We have seen that each set $\bE_A(N_i)$ is a neighborhood of 
$\alpha$ and we have pointed out earlier in this section that the 
two-dimensional convex body $L_A$ is stable. Hence 
(\ref{eq:barycenter}) shows that $\bE_A(N)$ contains a neighborhood of 
$\alpha$. This completes the proof.
\hspace*{\fill}$\square$\\
\par
The linear map $\bE_A|_{\cM_d}$ is open on the fiber 
$\bE_A|_{\cM_d}^{-1}(\alpha)$ of $\alpha\in L_A$ if $\alpha$ is a 
relative interior point of $L_A$ or a relative interior point of 
a facet of $L_A$. For a proof see \cite{Weis-cont}, Sec.~4.3, or 
\cite{Rodman-etal2016}, Sec.~3. Since $\dim(L_A)\leq 2$ we deduce 
from Thm.~\ref{thm:min-to-max} the following.
\begin{cor}\label{cor:strong-cont-open}
If $f_A^{-1}$ is strongly continuous at $\alpha\in L_A$ then $\bE_A|_{\cM_d}$ 
is open on the fiber $\bE_A|_{\cM_d}^{-1}(\alpha)$.
\end{cor} 
%
%
%
\section{Continuity Implies Strong Continuity}
\label{sec:MaxEnt-to-MinEnt}
\par
We prove that the continuity of $\rho_A^*$ implies the strong continuity 
of $f_A^{-1}$. This is the harder part compared to converse direction in
Sec.~\ref{sec:MinEnt-to-MaxEnt} because we now have to restrict the domain
from the state space $\cM_d$ to the pure states $\cM_d^1$ while keeping
the range $L_A$. A major argument will be a corollary of the
{\em strong stability} of the state space \cite{Shirokov2010}.
\par
It is well-known that the state space $\cM_d$ is stable. 
Indeed, Lemma~3 in \cite{Shirokov2006} proves that the map
\begin{equation}\label{eq:barycenter+Delta}\textstyle
\cM_d\times\cM_d\times[0,1]\to\cM_d,\quad
(\rho,\sigma,\lambda)\mapsto (1-\lambda)\rho+\lambda\sigma
\end{equation}
is open, which is equivalent to the stability of $\cM_d$ by Prop.~1.1 
in \cite{ClausingPapadopoulou1978}. To make an openness statement 
about $\bE_A|_{\cM_d^1}$ we have to restrict the left-hand side of 
(\ref{eq:barycenter+Delta}) from $\cM_d$ to $\cM_d^1$ while keeping 
the right-hand side. This restriction is indeed possible. The cost 
is the non-finiteness of the ensemble, see Rem.~1 in 
\cite{Shirokov2010}. The corresponding property of $\cM_d$ is called 
{\em strong stability} which, by definition, means that for all 
$k=1,\ldots,d$ the barycenter map from the discrete probability 
measures on $\{\rho\in\cM_d\mid\rk(\rho)\leq k\}$ to $\cM_d$ is open, 
see \cite{Shirokov2010}, Thm.~1. 
\par
Lemma~4 in \cite{Shirokov2010} serves for our purposes:
Let $\{\pi_i,\rho_i\}_{i\in\bN}$ be a countable ensemble, that is 
$\rho_i\in\cM_d^1$, $\pi_i\geq 0$ for all $i\in\bN$ 
and $\sum_{i=1}^\infty\pi_i=1$. For an arbitrary sequence 
$\{\rho^n\}\subset\cM_d$ converging to the average 
$\sum_{i=1}^\infty\pi_i\rho_i$ there exists a sequence 
$\{\{\pi_i^n,\rho_i^n\}_{i\in\bN}\}_{n\in\bN}$ of countable 
ensembles such that 
\begin{equation}\label{eq:M-strong-stability}\textstyle
\begin{array}{ll}
(\forall n) & \pi_1^n\rho_1^n+\pi_2^n\rho_2^n+\cdots=\rho^n,\\
(\forall i) & \lim_{n\to\infty}\pi_i^n=\pi_i 
\quad\mbox{and}\quad
(\pi_i>0\,\implies\,\lim_{n\to\infty}\rho_i^n=\rho_i). 
\end{array}
\end{equation}
We use an immediate corollary of (\ref{eq:M-strong-stability})
which is as follows.
\begin{cor}\label{cor:strong-stability-useful}
Let $\rho\in\cM_d^1$ be a pure state, let $N\subset\cM_d^1$ be a 
neighborhood of $\rho$ in $\cM_d^1$ and let $\sigma\in\cM_d$ be 
an arbitrary state. For every $\lambda\in[0,1)$ and every
$\tilde{\lambda}>\lambda$ with $\tilde{\lambda}\leq 1$ the set 
$(1-\tilde{\lambda})N+\tilde{\lambda}\cM_d$ is a neighborhood of 
$(1-\lambda)\rho+\lambda\sigma$ in $\cM_d$.
\end{cor}
\par
The Bloch ball $\cM_2$ with $\rho=|0\rangle\langle 0|$, 
$\sigma=|1\rangle\langle 1|$, and $\lambda\leq\tfrac{1}{2}$
shows that the assumption $\tilde{\lambda}>\lambda$ of 
Coro.~\ref{cor:strong-stability-useful} can not be weakened 
to $\tilde{\lambda}\geq\lambda$. 
\par
Based on two chapters of the numerical range theory, the next 
lemma provides an extremal point argument.  Firstly, Thm.~2 in 
\cite{Corey-etal2013} states 
that for all $\alpha\in L_A$ and $x\in S\bC^d$ such that $\alpha=f_A(x)$ 
and for any neighborhood $U$ of $x$ in $S\bC^d$ there is a constant 
$\delta>0$ such that $\delta L_A + (1-\delta)\alpha\subset f_A(U)$
holds. Secondly, Lemma 3.2 in \cite{Leake-etal-LMA2014} proves that 
for $r>0$ and $x\in S\bC^d$ the set
$f_A(\{y\in S\bC^d\mid|y-x|<r\})$ is convex.
\begin{lem}\label{lem:extremal-arg}
Let $\rho\in\cM_d^1$ and let $N\subset\cM_d^1$ be a neighborhood of 
$\rho$ in $\cM_d^1$. There exists a neighborhood 
$\widetilde{N}\subset N$ of $\rho$ in $\cM_d^1$ such that 
$\bE_A(\widetilde{N})$ is convex. The set $\bE_A(\widetilde{N})$ is a 
neighborhood of $\bE_A(\rho)$ in $L_A$ if it contains all extremal 
points of $L_A$ in a neighborhood of $\bE_A(\rho)$ in $L_A$.
\end{lem}
{\em Proof:}
The continuity of the quotient map $\beta:S\bC^d\to\cM_d^1$, see 
Lemma~\ref{lem:beta}, shows that $N':=\beta^{-1}(N)\subset S\bC^d$ 
is a neighborhood of any point in $\beta^{-1}(\rho)$. Let $x$ be 
such a point. Lemma 3.2 in \cite{Leake-etal-LMA2014}, cited above, 
shows that for some neighborhood $N''\subset N'$ of $x$ the image 
$f_A(N'')$ is convex. The openness of $\beta$ shows that
$\widetilde{N}:=\beta(N'')$ is a neighborhood of $\rho$ in $\cM_d^1$. 
Moreover, $\bE_A(\widetilde{N})=f_A(N'')$ is convex and 
contains $\alpha:=\bE_A(\rho)=f_A(x)$. This proves the first 
assertion.
\par
Let us prove that $f_A(N'')$ is a neighborhood of $\alpha$ 
if it contains all extremal points of $L_A$ sufficiently close to 
$\alpha$. We can assume that $L_A$ has real dimension two and that
$\alpha$ is an extremal point of $L_A$. Otherwise 
Thm.~2 in \cite{Corey-etal2013}, cited above, shows that 
$f_A(N'')$ is a neighborhood of $\alpha$.
\par
Since $f_A(N'')$ is convex, it suffices to show that it
contains a neighborhood of $\alpha$ in $\partial L_A$. We consider a disk 
$D:=\{\alpha'\in\bC\mid|\alpha'-\alpha|<\epsilon\}$ of radius 
$\epsilon>0$ about $\alpha$ and one of the semi-arcs, denoted by 
$r_\epsilon$, on the curve $D\cap\partial L_A$ and starting at 
$\alpha$. If $r_\epsilon$ is a segment for some 
$\epsilon$ then by Thm.~2 in \cite{Corey-etal2013} $f_A(N'')$ contains 
a neighborhood of $\alpha$ in $r_\epsilon$. Otherwise $r_\epsilon$ 
has a sequence of extremal points of $L_A$ which converges 
to $\alpha$. By assumptions, $f_A(N'')$ includes all extremal points 
of $L_A$ which are sufficiently close to $\alpha$. Hence, the convex
set $f_A(N'')$ contains the segments between those extremal 
points and therefore a neighborhood of $\alpha$ in $r_\epsilon$. 
Together with the analogous statement about the other semi-arc
we have shown that $f_A(N'')$ contains a neighborhood of $\alpha$.
\hspace*{\fill}$\square$\\
\par
We are ready to prove a main result of this paper.
\begin{thm}\label{thm:converse}
Let $\alpha\in L_A$. If $\bE_A|_{\cM_d}$ is open at a relative interior 
point of $\bE_A|_{\cM_d}^{-1}(\alpha)$ then $f_A^{-1}$ is 
strongly continuous at $\alpha$.
\end{thm}
{\em Proof:}
Let $\rho$ be a relative interior point of the fiber 
$F:=\bE_A|_{\cM_d}^{-1}(\alpha)$ such that $\bE_A|_{\cM_d}$ is open 
at $\rho$. We have to prove that $f_A$ is open at every point of
$f_A^{-1}(\alpha)$. Thus, using Lemma~\ref{lem:cont-m-m1-cs}, it 
suffices to prove that for all pure states $\sigma\in F$ the map 
$\bE_A|_{\cM_d^1}$ is open at $\sigma$.
\par
Let $N\subset\cM_d^1$ be a neighborhood of $\sigma$ in the set of
pure states $\cM_d^1$. By Lemma~\ref{lem:extremal-arg} there exists 
a neighborhood $N'\subset N$ of $\sigma$ in $\cM_d^1$ such that 
$\bE_A(N')$ is a neighborhood of $\alpha$ in $L_A$ provided that
it contains all extremal points of $L_A$ close to $\alpha$, which
we shall prove now.
\par
Let $\tau\in F$ and $\lambda\in(0,1)$ such that 
$\rho=(1-\lambda)\sigma+\lambda\tau$, and let $\tilde{\lambda}>\lambda$ 
with $\tilde{\lambda}<1$. Then Coro.~\ref{cor:strong-stability-useful}
proves that
\[\textstyle
N'':=(1-\tilde{\lambda})N'+\tilde{\lambda}\cM_d
\]
is a neighborhood of $\rho$ in $\cM_d$. By assumptions, 
$\bE_A|_{\cM_d}$ is open at $\rho$, so
\begin{equation}\label{eq:use-strong}\textstyle
\bE_A(N'')=(1-\tilde{\lambda})\bE_A(N')+\tilde{\lambda} L_A
\end{equation}
is a neighborhood of $\alpha$ in $L_A$. The definition of extremal
points and (\ref{eq:use-strong}) show that every extremal point of 
$L_A$ which lies in $\bE_A(N'')$ must lie in $\bE_A(N')$.
\hspace*{\fill}$\square$\\
\par
Thm.~\ref{thm:converse} 
and Coro.~\ref{cor:strong-cont-open} prove the following.
\begin{cor}\label{cor:major}
Let $\alpha\in L_A$. Then $\bE_A|_{\cM_d}$ is open at a relative 
interior point of $\bE_A|_{\cM_d}^{-1}(\alpha)$ if and only if
$f_A^{-1}$ is strongly continuous at $\alpha$. In this case
$\bE_A|_{\cM_d}$ is open on $\bE_A|_{\cM_d}^{-1}(\alpha)$.
\end{cor}
\par
We are ready to prove Theorem~\ref{thm1}.\\
\par
{\em Proof of Theorem~\ref{thm1}:}
As we have recalled in Rem.~\ref{rem1}(2), the inference state 
$\rho_A^*(\alpha)$ lies in the relative interior 
of $\bE_A|_{\cM_d}^{-1}(\alpha)$ for all $\alpha\in L_A$. Therefore 
the claim follows from (\ref{eqiv-cont-open}) and Coro.~\ref{cor:major}.
\hspace*{\fill}$\square$\\
%
%
%
%
%
\section{Continuity in Terms of Eigenvalue Functions}
\label{sec:rarity}
\par
We derive a continuity condition of the maximum-entropy 
inference from the theory of the numerical range map 
\cite{Leake-etal-LMA2014}. This shows that 
discontinuities of the maximum-entropy inference are the exception.
\par
For all $\theta\in\bR$ the hermitian matrix 
\[
\Re(e^{-\ii\theta}A)=\cos(\theta)\Re(A)+\sin(\theta)\Im(A)
\]
has an orthogonal basis of eigenvectors $\{x_k(\theta)\}_{k=1}^d$ 
and eigenvalues $\{\lambda_k(\theta)\}_{k=1}^d$ which depend real
analytically on $\theta$ \cite{Rellich1954}. Further, we define 
curves for $k=1,\ldots,d$,
\begin{equation}\label{eq:joswig-straub}
z_k(\theta)
:=e^{\ii\theta}(\lambda_k(\theta)+\ii\lambda_k'(\theta)),
\qquad \theta\in\bR,
\end{equation}
where $\lambda_k'$ is the derivative of $\lambda_k$ with respect 
to $\theta$. We remark that the union of these curves is a plane 
algebraic curve \cite{JoswigStraub1998} whose convex hull is the 
numerical range \cite{Kippenhahn1951}. 
\par
An eigenvalue function $\lambda_k$ is said \cite{Leake-etal-LMA2014}
to {\em correspond} to $\alpha\in W(A)$ at $\theta\in\bR$ if 
$z_k(\theta)=\alpha$. Notice by (\ref{eq:joswig-straub}), all 
eigenvalue functions corresponding to $\alpha\in W(A)$ at 
$\theta\in\bR$ have the same value and the same derivative at $\theta$.
We denote the support line of $W(A)$ with outward pointing normal 
vector $-e^{\ii\theta}$ by $\ell_\theta$. The following 
Fact~\ref{fac:leake}(1) is proved in Thm.~2.1(1) in 
\cite{Leake-etal-LMA2014}. Part (2) follows from Thm.~2 in 
\cite{Corey-etal2013}.
\begin{fac}\label{fac:leake}
\begin{enumerate}
\item
Let $\theta\in\bR$ and let $\alpha\in W(A)\cap\ell_\theta$ be an 
extremal point of $W(A)$. Then $f_A^{-1}$ is strongly continuous at 
$\alpha$ if and only if the eigenvalue functions corresponding to 
$\alpha$ at $\theta$ are mutually equal.
\item
Condition {\rm (1)} is for all points of $W(A)$ decisive, because 
$f_A^{-1}$ is strongly continuous at relative interior points of 
$W(A)$ and at relative interior points of facets. 
\end{enumerate}
\end{fac}
\par
An example and two corollaries will illustrate the use of 
Fact~\ref{fac:leake}.
\begin{exa}\label{ex:staffelberg}
A discontinuity of the maximum-entropy inference $\rho_A^*$ is known
\cite{WK,Chen-etal2014,Rodman-etal2016}
for 
\[
A:=\left[\begin{array}{ccc} 0 & 2 \\ 0 & 0 \end{array}
\right]\oplus [\,1\,],
\]
the direct sum denoting a block-diagonal matrix in $M_3$. The numerical 
range $W(A)$ is the unit disk in $\bC$ where $\rho_A^*$ is 
discontinuous at $1$. 
\par
Let us derive this discontinuity with new methods. The real 
part of the matrix $e^{-\ii\theta}A$ is
\[
\Re(e^{-\ii\theta}A)
=(\cos(\theta)\sigma_1+\sin(\theta)\sigma_2)
\oplus\cos(\theta),
\qquad\theta\in\bR,
\]
for Pauli matrices
\[
\sigma_1:=\left[\begin{array}{rr}0 & 1\\1 & 0\end{array}\right]
\quad\mbox{and}\quad
\sigma_2:=\left[\begin{array}{rr}0 & -\ii\\\ii & 0\end{array}\right].
\]
The eigenvalue functions are 
$\lambda_1(\theta)=1$, $\lambda_2(\theta)=-1$ and
$\lambda_3(\theta)=\cos(\theta)$ while
$z_1(\theta)=e^{\ii\theta}$, $z_2(\theta)=-e^{\ii\theta}$ and 
$z_3(\theta)=1$ holds. The eigenvalue functions corresponding 
to $1$ at $\pi$ are $\lambda_2$ and $\lambda_3$ and $1\in\ell_\pi$
holds. 
Since $\lambda_2\neq\lambda_3$, Fact~\ref{fac:leake} proves that 
$f_A^{-1}$ is not strongly continuous at $1$. Thus, 
Theorem~\ref{thm1} proves the discontinuity of $\rho_A^*$ at $1$. 
\end{exa}
\par
Finally we deduce, following \cite{Leake-etal-LMA2014}, that a 
discontinuity of the maximum-entropy inference is the exception.
The eigenvalue functions $\lambda_k$ extend analytically to a 
neighborhood of $\bR$ in $\bC$ and therefore, see for example 
\cite{Rudin}, Thm.~10.18, distinct eigenvalue functions can only 
coincide at finitely many exceptional values of $\theta\in[0,2\pi)$. 
Thus Fact~\ref{fac:leake}, Coro.~\ref{cor:major} and Thm.~\ref{thm1}
show the following.
\begin{cor}~
\begin{enumerate}
\item 
For all except possibly finitely many points $\alpha$ of $L_A$
the map $\bE_A|_{\cM_d}$ is open on $\bE_A|_{\cM_d}^{-1}(\alpha)$.
\item
For all except possibly finitely many points $\alpha$ of $L_A$
the maximum-entropy inference $\rho_A^*$ is continuous at $\alpha$.
\end{enumerate}
\end{cor}
\par
To make a statement about exceptionality of discontinuities 
in terms of observables we refrain from the global assumption 
that a matrix $A\in M_d$ is chosen. For all $d\in\bN$ the von
Neumann-Wigner non-crossing rule \cite{vonNeumannWigner1929} in 
the formulation of \cite{Gutkin-etal2004}, Prop.~4.9, states 
that the set of matrices $A\in M_d$ such that 
for any $(s,t)\in\bR^2\setminus\{0\}$ the hermitian matrix
$s \Re(A) + t \Im(A)$ has simple eigenvalues is open and dense
in $M_d$. Thus Fact~\ref{fac:leake}, Coro.~\ref{cor:major} 
and Thm.~\ref{thm1} show the following.
\begin{cor}
Let $d\in\bN$.
\begin{enumerate}
\item 
The set of matrices $A\in M_d$ where $\bE_A|_{\cM_d}$ is open
is open and dense in $M_d$.
\item
The set of matrices $A\in M_d$ where $\rho_A^*$ is continuous
is open and dense in $M_d$.
\end{enumerate}
\end{cor}
%
%
%
\section{Independence of the Prior State}
\label{sec:prior}
\par
The {\em MaxEnt} inference method 
\cite{ShoreJohnson1980,Skilling89,CatichaGiffin2006,Ali-etal2012} is 
an updating rule from a prior state to an inference state, given new 
information in terms of expected values. We show that the continuity
of the MaxEnt inference function does not depend on the prior state.
\par
As we have seen in Sec.~\ref{sec:overview} the set of expected 
values of two observables is the convex support $L_A$ which refers 
to a matrix $A\in M_d$, $d\in\bN$. By definition, the prior 
state $\rho\in\cM_d$ is assumed to be a positive definite matrix. 
The {\em MaxEnt} inference function, with respect to $A$ and $\rho$,
is 
\[
\Psi_{A,\rho}:\quad L_A\to\cM_d,\quad
\alpha\mapsto{\rm argmin}
\{S(\sigma,\rho)\mid\sigma\in\bE_A|_{\cM_d}^{-1}(\alpha)\}.
\]
This is a well-defined single-valued function \cite{Weis-cont}.
Here, the {\em Umegaki relative entropy} $S:\cM_d\times\cM_d\to[0,\infty]$ 
is an asymmetric distance which is zero only for equal arguments. By
definition, for states $\rho\in\cM_d$ of maximal rank holds
$S(\sigma,\rho)=\tr\sigma(\log(\sigma)-\log(\rho))$. Notice 
$S(\sigma,\id/d)=\log(d)-S(\sigma)$ for all $\sigma\in\cM_d$ where 
$S(\sigma)$ is the von Neumann entropy. So, $\Psi_{A,\id/d}=\rho_A^*$ 
is the maximum-entropy inference. Here $\id$ denotes the $d\times d$
identity matrix.
\par
The question whether the continuity of $\Psi_{A,\rho}$ depends on 
$\rho$ was asked in \cite{Weis-cont}, Rem.~5.9. For two observables
the answer is negative:
\begin{thm}\label{thm:MaxEnt}
All maps in the set 
$\{\Psi_{A,\rho}\mid\rho\mbox{\rm~is a prior state }\}$ 
have the same points of discontinuity in $L_A$.
\end{thm}
{\em Proof:}
Since the function $\sigma\mapsto S(\sigma,\rho)$ is continuous for 
each positive definite prior state $\rho\in\cM_d$, the continuity of 
$\Psi_{A,\rho}$ at $\alpha\in L_A$ is equivalent to the openness of 
$\bE_A|_{\cM_d}$ at $\Psi_{A,\rho}(\alpha)$, see \cite{Weis-cont}, 
Thm.~4.9. In addition, Lemma~5.8 in \cite{Weis-cont} proves that 
$\Psi_{A,\rho}(\alpha)$ lies in the relative interior of the fiber 
$\bE_A|_{\cM_d}^{-1}(\alpha)$. Now
Coro.~\ref{cor:major} proves the claim.
\hspace*{\fill}$\square$\\
%
%
%
\bibliographystyle{plain}

\begin{thebibliography}{10}

\bibitem{Alfsen-Shultz} E.\,M.~Alfsen, F.\,W.~Shultz (2001)
{\em State Spaces of Operator Algebras:\ Basic Theory, Orientations,
and C*-Products}, Springer-Verlag

\bibitem{Ali-etal2012} S.\,A.~Ali, C.~Cafaro, A.~Giffin, C.~Lupo, 
S.~Mancini (2012)
{\em On a differential geometric viewpoint of Jaynes' MaxEnt method 
and its quantum extension}, 
AIP Conf.\ Proc.\ 1443 120-128

\bibitem{Barndorff} O.~Barndorff-Nielsen (1978)
{\em Information and Exponential Families in Statistical Theory},
John Wiley \& Sons, New York

\bibitem{BZ} I.\ Bengtsson, K.\ \.Zyczkowski (2006)
{\em Geometry of Quantum States}, Cambridge University Press

\bibitem{BerberianOrland1967} S.\,K.~Berberian, G.\,H.~Orland (1967)
{\em On the closure of the numerical range of an operator},
Proc Amer Math Soc 18(3) 499-503

\bibitem{Buzek-etal1999} V.~Bu\v zek, G.~Drobn\'y, R.~Derka, G.~Adam, 
H.~Wiedemann (1999)
{\em Quantum state reconstruction from incomplete data},
Chaos, Solitons \& Fractals 10 981--1074

\bibitem{CatichaGiffin2006} A.~Caticha, A.~Giffin (2006)
{\em Updating Probabilities}, 
AIP Conf.\ Proc.\ 872 31--42

\bibitem{Chen-etal2014} J.~Chen, Z.~Ji, C.-K.~Li, Y.-T.~Poon, Y.~Shen,
N.~Yu, B.~Zeng, D.~Zhou (2015)
{\em Discontinuity of maximum entropy inference and quantum phase transitions},
\texttt{arXiv:1406.5046v2 [quant-ph]} 

\bibitem{ClausingPapadopoulou1978} A.~Clausing, S.~Papadopoulou (1978)
{\em Stable convex sets and extremal operators},
Mathematische Annalen 231 193--203

\bibitem{Corey-etal2013} D.~Corey, C.\,R.~Johnson, R.~Kirk, B.~Lins,
I.~Spitkovsky (2013)
{\em Continuity properties of vectors realizing points in the classical
field of values},
Linear and Multilinear Algebra 61(10) 1329--1338

\bibitem{FuchsVanDeGraaf1999} C.\,A.\ Fuchs, J.\ Van De Graaf (1999)
{\em Cryptographic distinguishability measures for quantum-mechanical states},
IEEE Trans Inf Theory 45(4) 1216--1227

\bibitem{Grzaslewicz} R.\ Grzaslewicz (1997) 
{\em Extreme continuous function property},
Acta Mathematica Hungarica 74(1--2) 93--99

\bibitem{Gutkin-etal2004} E.~Gutkin, E.\,A.~Jonckheere, M.~Karow (2004)
{\em Convexity of the joint numerical range: topological and 
differential geometric viewpoints},
Linear Algebra and its Applications 376 143--171

\bibitem{HornJohnson1991} R.\,A.~Horn, C.\,R.~Johnson (1991)
{\em Topics in Matrix Analysis}, Cambridge University Press, Cambridge

\bibitem{Ingarden-etal1997} R.\,S.~Ingarden, A.~Kossakowski, M.~Ohya (1997)
{\em Information Dynamics and Open Systems}, Kluwer Academic Publishers Group

\bibitem{Jaynes1957} E.\,T.~Jaynes (1957)
{\em Information theory and statistical mechanics.},
Phys Rev 106 620--630 and 108 171--190

\bibitem{JoswigStraub1998} M.~Joswig, B.~Straub (1998)
{\em On the numerical range map}, 
Journal of the Australian Mathematical Society 65 267--283

\bibitem{Kato-etal2015} K.~Kato, F.~Furrer, M.~Murao (2015)
{\em Equivalence of topological entanglement entropy and 
irreducible correlation and relationship to secret sharing},
\texttt{arXiv:1505.01917 [quant-ph]} 

\bibitem{Kippenhahn1951} R.~Kippenhahn (1951)
{\em \"Uber den Wertevorrat einer Matrix},
Mathematische Nachrichten 6(3--4) 193--228

\bibitem{Leake-etal-OAM2014} T.~Leake, B.~Lins, I.\,M.~Spitkovsky (2014)
{\em Pre-images of boundary points of the numerical range},
Operators and Matrices 8(3) 699--724

\bibitem{Leake-etal-LMA2014} T.~Leake, B.~Lins, I.\,M.~Spitkovsky (2014)
{\em Inverse continuity on the boundary of the numerical range},
Linear and Multilinear Algebra 62 1335--1345

\bibitem{LiPoon2000} C.-K.~Li, Y.-T.~Poon (2000)
{\em Convexity of the joint numerical range},
SIAM J Matrix Anal A 21(2) 668--678

\bibitem{Linden-etal2002} N.~Linden, S.~Popescu, W.~Wootters (2002)
{\em Almost every pure state of three qubits is completely determined 
by its two-particle reduced density matrices},
Physical Review Letters 89(20) 207901

\bibitem{LinsParihar2015} B.~Lins, P.~Parihar (2015)
{\em Continuous selections of the inverse numerical range map},
Linear and Multilinear Algebra 1--13

\bibitem{Liu-etal2014} Y.~Liu, B.~Zeng, D.\,L.~Zhou (2014)
{\em Irreducible many-body correlations in topologically ordered systems},
\texttt{arXiv:1402.4245 [quant-ph]} 

\bibitem{vonNeumannWigner1929} J.~von~Neumann, E.\,P.~Wigner (1929)
{\em \"Uber das Verhalten von Eigenwerten bei adiabatischen Prozessen},
Physikalische Zeitschrift 30 467--470

\bibitem{NC} M.\,A.\ Nielsen, I.\,L.\ Chuang (2000)
{\em Quantum Computation and Quantum Information},
Cambridge University Press

\bibitem{Papadopoulou1977} S.~Papadopoulou (1977)
{\em On the geometry of stable compact convex sets},
Math Ann 229 193--200

\bibitem{ProtasovShirokov2009} V.\,Y.~Protasov, M.\,E.~Shirokov (2009) 
{\em Generalized compactness in linear spaces and its applications},
Sbornik:\ Mathematics 200(5) 697--722

\bibitem{Rellich1954} F.~Rellich (1954)
{\em Perturbation Theory of Eigenvalue Problems},
Research in the Field of Perturbation Theory and Linear Operators,
Technical Report No.~1, 
Courant Institute of Mathematical Sciences, New York University

\bibitem{Rockafellar} R.\,T.~Rockafellar (1972)
{\em Convex Analysis},
Princeton University Press

\bibitem{Rodman-etal2016} L.~Rodman, I.\,M.~Spitkovsky, 
A.~Szko{\l}a, S.~Weis (2016)
{\em Continuity of the maximum-entropy inference:\
Convex geometry and numerical ranges approach},
Journal of Mathematical Physics 57, 015204

\bibitem{Rudin} W.~Rudin (1987)
{\em Real and Complex Analysis}, 3rd ed., McGraw-Hill

\bibitem{Shirokov2006} M.\,E.~Shirokov (2006)
{\em The Holevo capacity of infinite dimensional channels and the
additivity problem}, Commun Math Phys 262 137--159

\bibitem{Shirokov2010} M.\,E.~Shirokov (2010)
{\em Continuity of the von Neumann entropy},
Commun Math Phys 296(3) 625--654

\bibitem{Shirokov2012} M.\,E.~Shirokov (2012)
{\em Stability of convex sets and applications},
Izvestiya: Mathematics 76(4) 840--856

\bibitem{ShoreJohnson1980} J.\,E.~Shore, R.\,W.~Johnson (1980) 
{\em Axiomatic derivation of the principle of maximum entropy 
and the principle of minimum cross-entropy}, 
IEEE Trans.\ Inf.\ Theory 26 26--37; 
Correction (1983) ibid.\ 29, 942--943

\bibitem{Skilling89} J.~Skilling (1989)
{\em Classic maximum entropy},
Maximum Entropy and Bayesian Methods, Springer 45--52.

\bibitem{Streater2011} R.\,F.~Streater (2011)
{\em Proof of a modified Jaynes's estimation theory},
Open Systems \& Information Dynamics 18(2) 223--233

\bibitem{SwingleKim2014} B.~Swingle, I.\,H.~Kim (2014)
{\em Reconstructing quantum states from local data},
Physical Review Letters 113(26) 2014

\bibitem{Uhlmann1976} A.~Uhlmann (1976)
{\em The ``transition probability'' in the state space of a *-algebra},
Reports on Mathematical Physics 9(2) 273--279

\bibitem{Uhlmann2010} A.~Uhlmann (2010)
{\em Roofs and convexity},
Entropy 12(7) 1799--1832

\bibitem{Wehrl1978} A.~Wehrl (1978) 
{\em General properties of entropy},
Reviews of Modern Physics 50 221--260

\bibitem{Weis-topo} S.~Weis (2014)
{\em Information topologies on non-commutative state spaces},
Journal of Convex Analysis 21(2) 339--399

\bibitem{Weis-cont} S.~Weis (2014)
{\em Continuity of the maximum-entropy inference},
Communications in Mathematical Physics 330(3) 1263--1292

\bibitem{Weis-GSI14} S.~Weis (2014) 
{\em The MaxEnt extension of a quantum Gibbs family, convex geometry 
and geodesics},
AIP Conf.\ Proc.\ 1641 173--180.

\bibitem{WK} S.~Weis, A.~Knauf (2012)
{\em Entropy distance:\ New quantum phenomena},
J Math Phys 53(10) 102206

\bibitem{Wichmann1963} E.\,H.~Wichmann (1963)
{\em Density matrices arising from incomplete measurements},
Journal of Mathematical Physics 4(7) 884--896

\bibitem{Zhou2008} D.~Zhou (2008)
{\em Irreducible multiparty correlations in quantum states without maximal rank},
Physical Review Letters 101(18) 180505

\end{thebibliography}

\vspace{.5in}
Stephan Weis\\
Departamento de Matemática\\
Instituto de Matemática, Estatística e Computação Científica\\
Universidade Estadual de Campinas\\
Campinas - SP - 13083-859\\
Brazil\\
e-mail: maths@stephan-weis.info
\end{document}